\documentclass{article}  
\usepackage{breckenridge}
\usepackage{graphicx}
\frompage{000} \topage{000}                                              
\def\rightmark{Freezeout of  Resonances}
\def\leftmark{E.V.Shuryak}
\newcommand {\be}{\begin{eqnarray}}
\newcommand {\ee}{\end{eqnarray}}

\title{Freezeout of resonances and nuclear fragments  at RHIC} 
\authors{E.V.Shuryak \\
Department of Physics and Astronomy, \\ 
SUNY at Stony Brook, NY 11794, USA 
}
 
\abstract{We quantify the conditions at which ``composites'', the
 resonances and bound states $d,He^3$ 
 are produced at RHIC. Using Hubble-like model for late stages,
  one can analytically solve the rate equations
 and also
calculate the relevant optical depth factors. We calculate also the
modification of $\rho$ masse and width,  and predict a radiacal
shape change of $\sigma$.
}

 
\begin{document}
 
\maketitle
\setcounter{page}{1}

\section{Introduction}\label{intro}
This talks is based on two papers, one with G.Brown \cite{Shuryak:2002kd}
and one ongoing work with P.Kolb \cite{KS_03}, who will  report a part of it
related with late time evolution  
 here. Their common goal is to understand what are the
conditions which determine the timing of production of 
$observable$ resonances,
and estimate  
at what conditions this happens. We have emphasized
$observable$ above because there are many resonances produced in the system
but unobserved in the final invariant mass spectra because of
re-scattering
of their decay products. The development of simple analytic model
of the kinetics of resonance production/absorption, as well as evaluation
of the ``optical depth'' integral.

An old but still important idea is ``matter modification'' of hadrons,
providing the experimental test of the conditions in question. As such
we discuss modification of the mass and shape of two classic $\pi\pi$ 
resonances, $\rho$ and $\sigma$. Recent STAR observations of these effects
 are reproduced (for $\rho$) and predicted (for $\sigma$).

The issue of production of nuclear
fragments is also an old one, at RHIC reduced to
 $d,He^3$ and their antiparticles.  In literature those are studied
either by statistical or coalescence models, which left open many
important issues. First of all, like resonances the $observable$ 
 fragments must escape all interactions
with any particles in order to survive at the end. The second point
is that this process is production-rate-limited, thus it can lead
to non-thermal quasi-equilibria. Furthermore, new element 
is the consistent evaluation of the 2-to-1 production rate itself
in a recent work by Ioffe et al \cite{Ioffe:2003uf}.

\section{The optical depth factor}\label{sec_depth}  
Let me start with the simplest pedagogical points about the 
observability of resonances and fragments, or ``composites'' as we may
call them collectively, for brevity.

In the next section we will discuss rate equations which can be solved and
determine the number of composites $N(t)$ at time $t$.
The ``observability condition'' of a resonance can be written as
\be \nu_{visible}(t)=\Gamma N(t) exp\left(-\int_t^{\infty}\nu(t')dt'\right)\ee
where the the l.h.s. is the production rate of visible resonances,
$\Gamma$
is the resonance decay width and the exponent is the optical depth
factor
containing integrated 
$\nu(t)$, the combined scattering rate for all decay products.
The $N(t)$ decreases with time due to
expansion and cooling, while the exponent changes from 0 at early
time to 1 at late times: so the product naturally has a maximum at
 the time $t_m$ such that
\be 
{1\over N(t_m)}{dN(t_m)\over dt}+\nu(t_m)=0
\ee
This condition means that for observable resonances the freezeout
condition
is different from that for stable particles and reads:
{\em the rate of their number change is equal to the absorption
  rate of all the decay products}. 
For example, for $\rho$ and $\sigma$ we should not know their scattering
rates but just that of two pions.
Since for short-lived $\rho$ and $\sigma$ the first factor is
close
to overall expansion rate of matter at late time which
follows from Hubble-like late-time regime $d \log
N(t_m)/dt\approx 3/t$, and the second is  the same,
we conclude that  `visible''  $\rho$ and
$\sigma$
are produced at the same time.  

The formation rate for a fragment made of $A$ nucleons is
made by some coalescence, and such rate is obviously proportional
to nucleon density to that power, $\sim(n_N(t))^A$.
After it is produced,
 however, it still
has very small probability to survive.
 Assuming that the
destruction rate for a fragment  $\nu_A\approx A\nu_N$, where $\nu_N$
is a scattering rate for one nucleon,   
one finds that for $A$-fragment the time distribution is approximately
the $A$-th power of the same universal function
\be n_{fragments}(t)\sim \left[ n_N(t)
  exp\left(-\int_t^{\infty}\nu_N(t')dt'\right) \right]^A\ee
So, {\em the maximum 
of production of any visible fragment happens at the same time for all $A$}.
Furthermore,
  the width of the distribution over production time  
decreases as $A$ grows, as $1/\sqrt{A}$. 

\section{Solving the rate equations}\label{sec_rate_eqn}

The equations themselves describing dynamics of resonances are well
known, generically they
contain  the sink (the decay) and 
the source terms 
\be {\partial n(t,\vec r)\over \partial t}=-\Gamma n(t,\vec
r)+S(n_i)\ee
(where for expanding source the time should be understood as proper time
in the rest frame of all volume elements.)
 In many papers in literature
(e.g. \cite{Markert:2002rw}) the source term is ignored citing
 ``instantaneous hadronization'', but (especially 
for resonances we consider) it is not true: in fact
the primary
generation of resonances die out long before the ``observable'' ones
are born.

\begin{figure}[h!]
\begin{minipage}[c]{6.cm}
 \centering 
\includegraphics[width=4.cm]{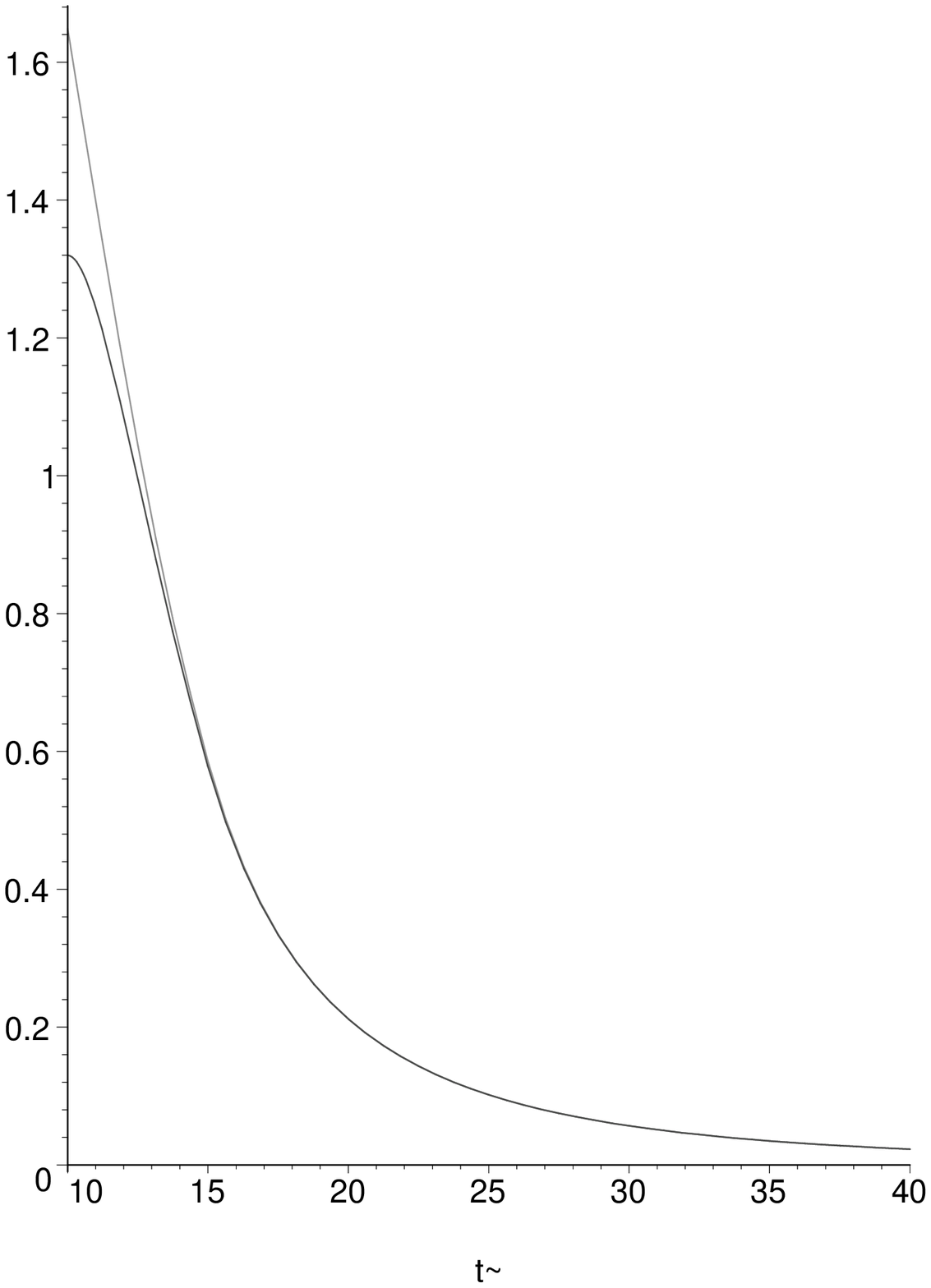}
 \end{minipage}
\begin{minipage}[c]{6.cm}
 \centering 
\includegraphics[width=4cm]{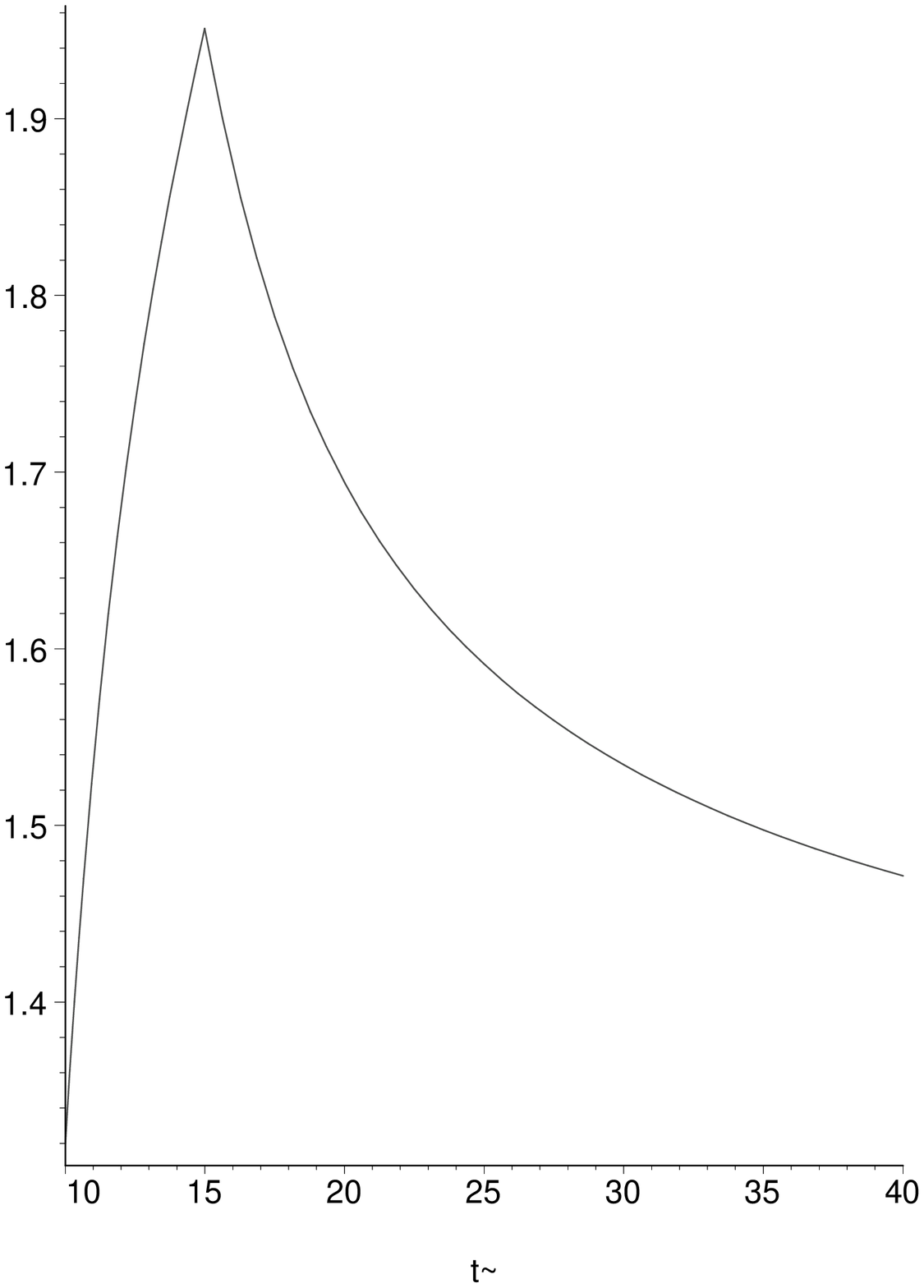}
\end{minipage}
\caption{
\label{fig_rho_kinetics}
(a) The time dependence of the $rho$-meson density,
starting from chemical equilibrium at $t=5$ and $10\, fm$. Plotting
its ratio to the pion density in (b) one observes
the transition from hydro to free streaming regimes
and
that the $\rho$ density decreases power-like rather
than exponentially, because
of the source term.
}
\end{figure}

 We use an approximate power fit of the source time dependence
$ \int d^3r S=\Gamma N_0 \left({t_0 \over t}\right)^P $
Its power can be related to fireball expansion. If
 the volume $V(t)\sim 1/n(t)\sim t^a$ 
 the integrated
source is proportional to $V(t)[n_\pi(t)]^N$ where $N$ is the
multiplicity in resonance decay
 ($N=2$ for $\sigma,\rho$). The pion number is
``chemically frozen'', $N_\pi=V(t)n_\pi(t)=const(t)$ from which we
conclude that the source term power is $P=(N-1)a$ (for $\sigma,\rho$
and other binary resonances $P=a$ ). An example of such equation solved
is shown in Fig.\ref{fig_rho_kinetics}.

\begin{figure}[h!]
\begin{minipage}[c]{6.cm}
 \centering 
{\includegraphics[width=4cm]{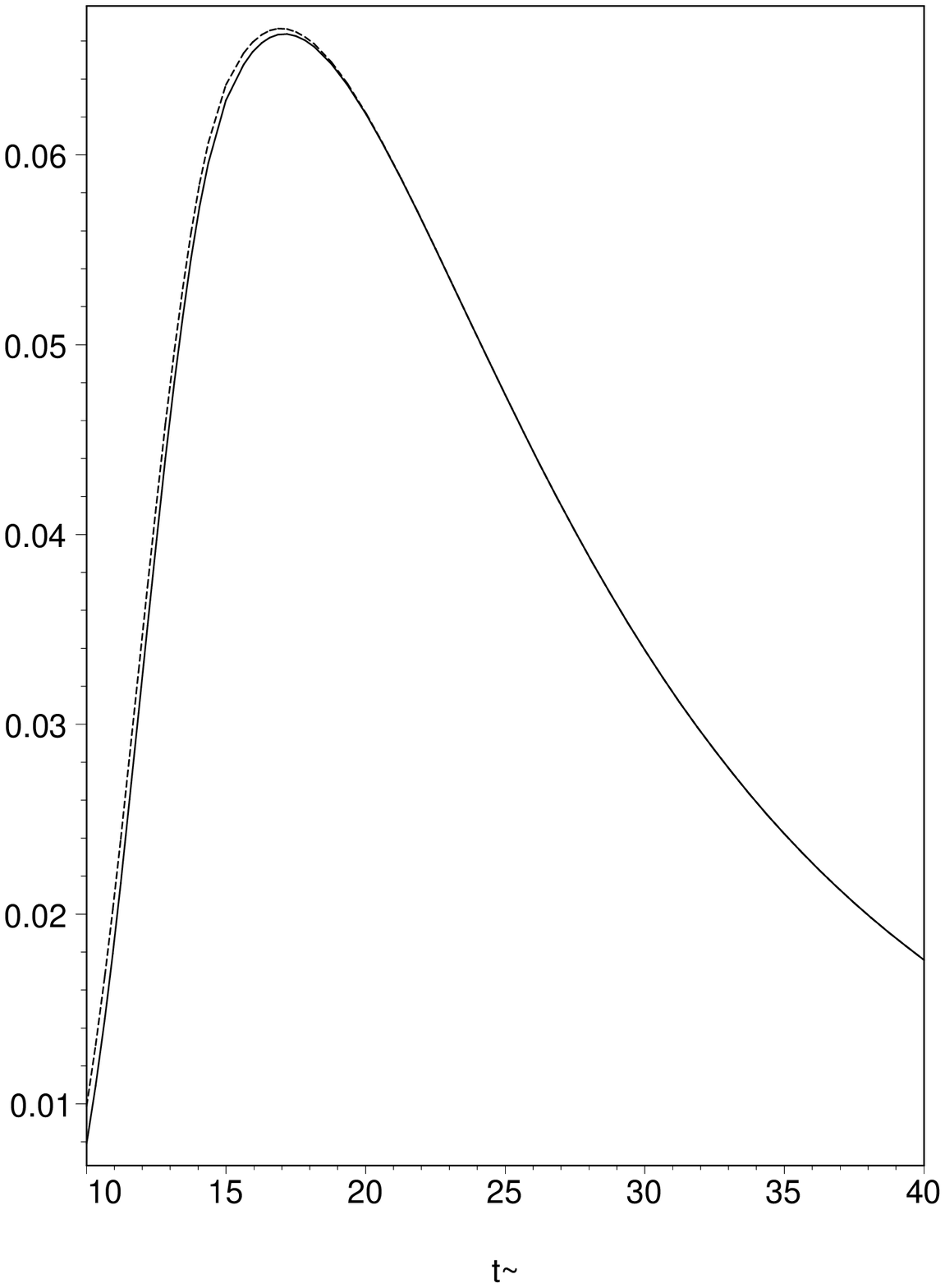}}
 \end{minipage}
\begin{minipage}[c]{6.cm}
 \centering 
\includegraphics[width=4cm]{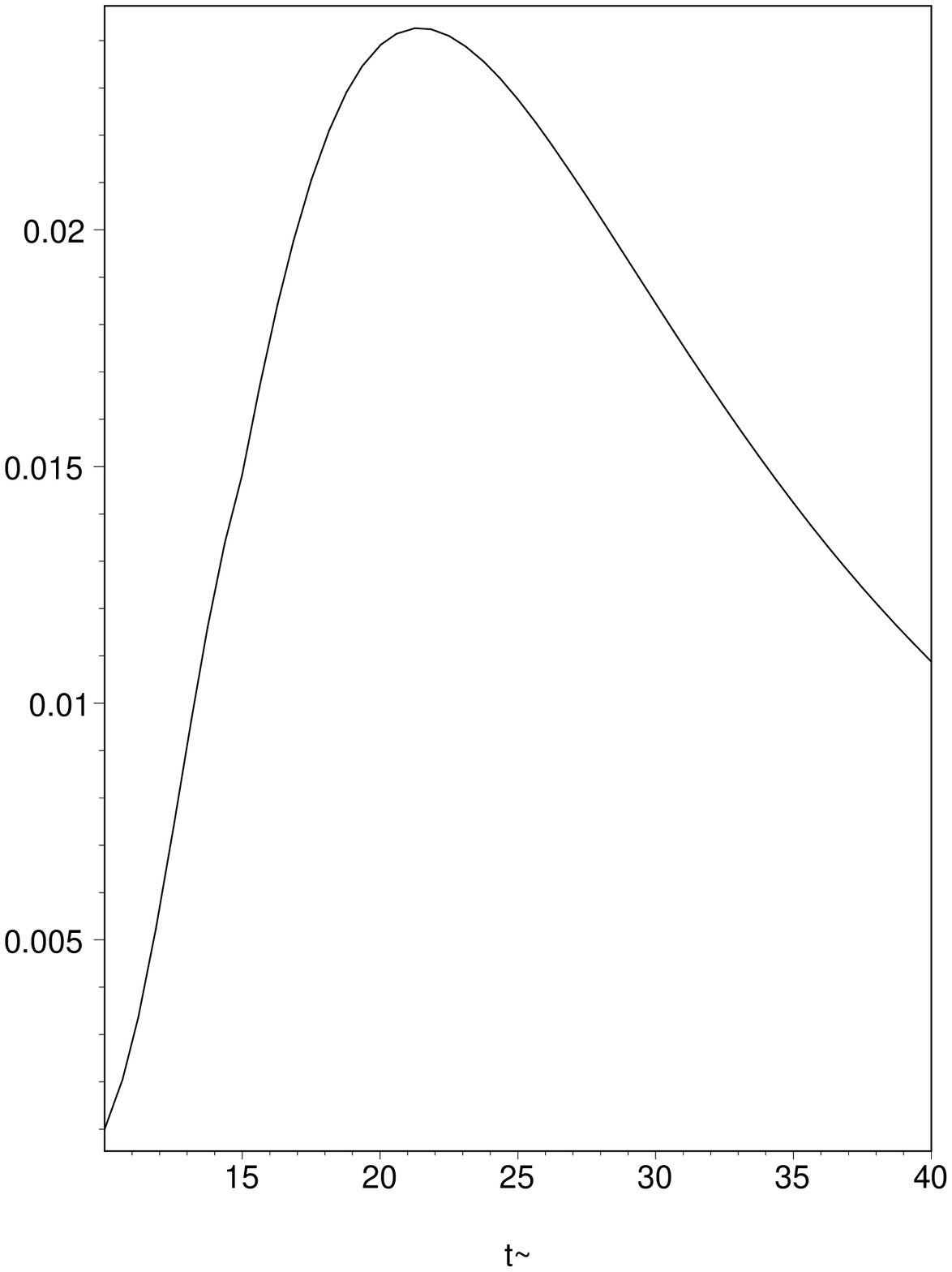}
\end{minipage}
\caption{
\label{fig_rho_d}
 The time dependence of the visible $rho$-meson (a) and deuteron (b)
production, for the $r=0$ point in central AuAu collisions at RHIC.
}
\end{figure}

We then evaluate the
 {\em optical factors} for pions and nucleons, using 
realistic
re-scattering rates, with chemically frozen composition, using papers
by Hung and myself \cite{Hung:1997du} and 
by Tomasik and Wiedemann \cite{Tomasik:2002rx}. Example for
 the final time distributions for {visible}
$\rho$ and $d$ is shown in Fig.\ref{fig_rho_d}. Note that
both distribution have maxima we discussed above,
and that the ``visible'' $d$ are indeed produced very late.
This is our main point: we are speaking about a very dilute matter,
after the freezeout of all the basic ingredients of the
 fireball.

\section{The resonance modification}
It has been argued
over the years that in matter the
resonances should be modified, with
 shifted
mass, increased width and even significantly changed shape. With
the very late stages of RHIC collisions relevant, we can now access
very dilute matter in which those effects must be calculable in the
lowest order of the density, providing a benchmark test to all such
discussions. In such case hadron modification is
expressed in terms of their {\em forward
scattering amplitude} $M_{ij}(t=0,s)$.
Note that the scattering amplitude is complex, and that this approach
gives
both the real and imaginary part of the dispersion law modification, also known as
the optical potential. 

There are two
major theoretical approaches to the issue discussed in literature,
 to be called an $s-channel$ and a $t-channel$ one.
The former approach assumes that the scattering amplitude is dominated by
s-channel resonances which are known to decay into the $i+j$  channel.
For most mesons such as
$\pi,\omega,\rho,K$  in a gas made of pions such calculation has been
made e.g. in \cite{Shu_pots} related with
such $\pi\rho$ resonances as $a_1$  or  $N\rho$ resonance $N^*(1520)$
\cite{CRW} for $\rho$ at SPS. Note that the signs of the effects are 
opposite in those examples, as seen from the following table:

\begin{table}[h!]
\caption{A set of resonances considered}
\centering
\begin{tabular}{lcrr}\hline
\it Name/Mass &\it Width &\it Branching & Mass Shift  \cr\hline
$a_1(1260)$ & 400 & 0.6 & -19 \cr
$a_2(1320)$ & 104 & 0.7 &  -15\cr
$K_1(1270)$ & 90 & 0.4 & +1.6 \cr
$K_2(1430)$ & 100 & 0.087 & -0.4 \cr
$N^*(1520)$ & 370 & 0.2 & 10 \cr
\hline
\end{tabular}
\label{table1a}
\end{table}

The majority of the particles in the matter
are Goldstone bosons $\pi,K,\eta$ which do not
interact at small momenta. 
However attraction between
 other particles is there.
Using a simple
expression for
the
 mass shift one gets
$ \delta m
_\rho^N \approx -28 MeV $
due to all
 $\bar B+B$. An additional shift 
$ \delta m_\rho^v \approx -10 MeV $
comes from scalar exchanges between $\rho$ and all other 
 vector mesons $\rho,\omega,K^*$. 
The main difference between the two mechanisms of the mass shift
discussed above is that the t-channel attraction 
is not associated with the broadening, while the
s-channel resonances increase the width by
 about 50 MeV. 
On the other hand, there is a ``kinematic'' effect working to the
opposite direction. The
negative mass shift discussed above automatically reduces the width,
both because of the reduced phase space and also due to the power of p
in the P-wave matrix element. The magnitude of this effect for
the predicted mass shift is
\be \delta \Gamma_\rho = 3{\delta m_\rho \over m_\rho}\Gamma_\rho \approx -50 \, MeV \ee
So, inside the accuracy these two effects cancel each other.

The invariant mass distributions in pp and mid-central AuAu
 of the $\pi^+\pi^-$ system, with a
transverse momentum cut
$0.2 < p_t < 0.9 \, GeV$ have been measured by STAR
 \cite{STAR_Fachini}, see also the C.Markert's talk here.
I would not have time here to discuss the shift in pp (see
 \cite{Shuryak:2002kd}), and I only comment that in AuAu the $\rho$ peak
 is
found to be shifted by additional $\sim -40\, MeV$ in mass, but  the
 width is the same. This agrees well with estimates above.

The same approach should of course be applied to many other
resonances. For more narrow resonances, like $K^*$, we expect smaller
shifts,
while for wider resonances like $\sigma$ we predicted a complete
change
of shape. At small freezeout $T\approx 100 \, MeV$ sigma 
was predicted to be
deformed
into a much more narrow structure at mass of about 400 MeV, see figures
in
\cite{Shuryak:2002kd}. 
Exactly such a peak has been seen by
STAR\footnote{I have seen it first at this workshop, on the day after
  my talk, shown by
  Gary Westfall in a plot of the balance function as a function of
  $q_{inv}$ with a peak he said ``nobody understands''. It was rather good
timing between the prediction and its 
experimental confirmation.}. Another 
confirmation
of very late freezeout and low $T$, 
the $\sigma/\rho$ ratio strongly grows toward central collisions.
 We are waiting for quantitative analysis of
these data with great interest.

\section{Fragment coalescence}

The issue of coalescence, such as $p+n\rightarrow d$,  was discussed in
many
papers over the years, and authors struggled with the question how
to calculate its rate. In particular, it is clear that when the level
crosses zero the wave function at the origin vanishes, and so the production
should do so too. And if the production rate is small compared to 
two other relevant rates, $\nu_{abs}$ and $dlog N/dt$, there is never
thermal equilibrium and one should not use statistical models. 

Significant progress has been made in recent paper by Ioffe et al
\cite{Ioffe:2003uf} who have pointed out how to use consistently the
in-matter
widths of all particles and obtain the production rate. We are now
incorporating it into the picture of expansion and the optical depths
discussed above, and hope to get quantitative results for $\bar d,d$ spectra
soon.

\section*{Acknowledgements}
My first thanks go to  the organizers, especially to the hard core from MSU,
who wonderfully managed to collect us  together year after
year in the right places.
 I should also thank my collaborators, G.Brown and P.Kolb, and also
 Patricia Fachini who made analysis of the STAR data for $\pi\pi$
 resonances and explained it to us.
Finally, my research is
supported by the DOE grant No. DE-FG02-88ER40388.

\vfill\eject
\end{document}